\documentclass[journal]{IEEEtran}

\usepackage{url}
\usepackage{amssymb,amsmath}
\usepackage[all]{xy}
\usepackage{bm}
\usepackage[lined,boxed,commentsnumbered]{algorithm2e}
\usepackage{subfigure}
\usepackage{graphicx,color,psfrag}
\usepackage{xspace}
\usepackage{url}
\usepackage{cite,xspace,xy,soul}
\usepackage{dsfont}
\usepackage{cases}
\usepackage{psfrag}
\usepackage{pdfsync}
\usepackage{soul}
\usepackage[update,prepend]{epstopdf}

\newcommand{\varspace}{\bm{d}, \bm{p}}

\newcommand{\sinr}{\mathsf{SINR}}
\newcommand{\usum}{U^{\rm{sum}}}
\newcommand{\LIN}{\mathsf{LIN}}
\newcommand{\LOG}{\mathsf{LOG}}
\newcommand{\DLOG}{\mathsf{DLOG}}

\newcommand{\lmax}{x^{\star}_{\mathrm{max}}}
\newcommand{\rmaxone}{\rho}

\newcommand{\journal}[1]{#1}

\begin{document}

\title{Data Offloading in Load Coupled Networks: \\ A Utility Maximization Framework}

\journal{
\author{Chin Keong Ho, Di Yuan, and~Sumei Sun%
\thanks{This paper is presented in part at the IEEE International Conference on Communications, June 2013.}
\thanks{C. K. Ho and S. Sun are with the Institute for Infocomm Research, A*STAR, 1 Fusionopolis Way, \#21-01 Connexis, Singapore 138632 (e-mail: \{hock, sunsm\}@i2r.a-star.edu.sg).}
\thanks{D. Yuan is with the Department of Science and Technology, Link{\"o}ping University, Sweden. (e-mail: di.yuan@liu.se)}
\thanks{The work of D. Yuan has been supported by the Swedish ELLIIT Excellence Center, CENIIT of Link{\"o}ping University, Sweden, and the European FP7 Marie Curie IAPP scheme with contract number 324515.}
}
}

\newtheorem{conjecture}{Conjecture}
\newtheorem{remark}{Remark}
\newtheorem{insight}{Insight}
\newtheorem{question}{Question}
\newtheorem{proposition}{Proposition}
\newtheorem{corollary}{Corollary}
\newtheorem{lemma}{Lemma}
\newtheorem{assumption}{Assumption}
\newtheorem{theorem}{Theorem}
\newtheorem{example}{Example}
\newtheorem{property}[theorem]{Property}

\newcommand{\myse}{\IEEEyessubnumber} 
\newcommand{\myses}{\myse\IEEEeqnarraynumspace} 

\newcommand{\set}[1]{\mathcal{#1}}

\newcommand{\bn}{\begin{enumerate}}
\newcommand{\en}{\end{enumerate}}

\newcommand{\bi}{\begin{itemize}}
\newcommand{\ei}{\end{itemize}}

\newcommand{\be}{\begin{IEEEeqnarray}{rCl}}
\newcommand{\ee}{\end{IEEEeqnarray}}

\newcommand{\benl}{\begin{IEEEeqnarray*}}
\newcommand{\eenl}{\end{IEEEeqnarray*}}

\newcommand{\bel}{\begin{IEEEeqnarray}}
\newcommand{\eel}{\end{IEEEeqnarray}}

\newcommand{\ben}{\begin{IEEEeqnarray*}{rCl}}
\newcommand{\een}{\end{IEEEeqnarray*}}

\newcommand{\barr}{\begin{array}}
\newcommand{\earr}{\end{array}}

\newenvironment{definition}[1][Definition:]{\begin{trivlist}
\item[\hskip \labelsep {\it #1}]}{\end{trivlist}}

\newcommand{\ud}{\mathrm{d}} 

\newcommand{\FigSize}{0.6}
\newcommand{\FigSizeSmall}{0.5}

\newcommand{\avesnr} {\bar{\gamma}} 
\newcommand{\snr} {\gamma} 

\newcommand{\re}[1]{(\ref{#1})}

\newcommand{\Pe} {P_{\mathrm {e}}} 

\newcommand{\goodgap}{%
\hspace{\subfigtopskip}%
\hspace{\subfigbottomskip}}

\newcommand{\dhat}[1]{\Hat{\Hat{#1}}} 
\newcommand{\that}[1]{\Hat{\Hat{\Hat{#1}}}} 
\newcommand{\dtilde}[1]{\Tilde{\Tilde{#1}}} 
\newcommand{\ttilde}[1]{\Tilde{\Tilde{\Tilde{#1}}}} 

\newcommand{\trace}[1]{\mathrm{tr}\{#1\}} 

\newcommand{\diag}{\mathop{\mathrm{diag}}}
\newcommand{\load}{x}
\newcommand{\loadv}{\bm{\load}}
\newcommand{\ymax}{Y}
\newcommand{\Pmax}{P_{\rm{max}}}

\maketitle


\begin{abstract}
We provide a general framework for the problem of data offloading in a heterogeneous wireless network, where some demand of cellular users is served by a {\em complementary network}. The complementary network is either a small-cell network that shares the same resources as the cellular network, or a WiFi network that uses orthogonal resources. For a given demand served in a cellular network, the load, or the level of resource usage, of each cell depends in a non-linear manner on the load of other cells due to the mutual coupling of interference seen by one another. With load coupling, we optimize the demand to be served in the cellular or the complementary networks, so as to maximize a utility function. We consider three representative utility functions that balance, to varying degrees, the revenue from serving the users vs the user fairness. We establish conditions for which the optimization problem has a feasible solution and is convex, and hence tractable to numerical computations. Finally, we propose a strategy with theoretical justification to constrain the load to some maximum value, as required for practical implementation. Numerical studies are conducted for both under-loaded and over-loaded networks.
\end{abstract}




%


\section{Introduction}

Fueled by multimedia mobile applications, the demand for mobile data is rising rapidly. Data traffic is also projected to grow at a compound annual growth rate of $78\%$ from 2011 to 2016 \cite{Cisco}.
In practice, cellular networks and the conventional infrastructure cannot grow as fast to match the increase in demand.
One promising solution currently considered by cellular operators is to employ data offloading, also known as mobile cellular traffic offloading \cite{LeRhLeChYi10,HaHuSr11}. In data offloading, the data of cellular users is intentionally delivered by complementary networks, namely small cells such as Picocells and Femtocells, or WiFi networks. This reduces the data demand on the regular cellular networks and hence eases traffic congestion. 

In a cellular network, frequency reuse is employed, and thus base stations using the same frequency band interfere with one another. We refer to the average level of resource usage in the time-frequency domain a cell as its load.
To optimize the  overall system performance, load balancing has to be performed across various networks in the context of data offloading \cite{song2007load}.
Due to the mutual coupling of the interference and the requirement to serve a specific demand for each cell, the load of a cell depends on the load of other cells. This leads to a non-linear coupling relation of the cells' loads, making analytical characterization of the load challenging. This motivates the use of new approaches and different theoretical tools to analyze and optimize the system performance.

Recently, an analytical signal-to-interference-and-noise-ratio (SINR) model that takes into account the load of each cell is employed \cite{SiFuFo09, MaKo10}, resulting in a non-linear load coupling equation for which theoretical analysis is obtained in \cite{SiominaYuan}.
This load coupling equation has also been shown to give a good approximation for more complicated load models in cellular systems that capture the dynamic nature of arrivals and service periods of data flows in the network, especially at high data arrival rates \cite{FehskeFettweisICC12}.
For example, the load obtained by the load coupling equation is within $10\%$ of the exact load obtained, if the normalized arrival rate of the data is more than $60\%$.

In this paper, we consider two separate scenarios in which the cellular network offloads to a complementary network. The complementary network is either a small cell or a WiFi network.
The small-cell network shares the network resources with the cellular network, whereas the WiFi network uses orthogonal network resources to that of the cellular network.
The performance of serving users in a particular network is measured by three representative types of utility functions, all of which are related to the network operator's revenue, but differ in the degree of accounting for user fairness. To model the inter-dependency of the load, we employ the load coupling equation in \cite{SiFuFo09, MaKo10, SiominaYuan, FehskeFettweisICC12}.

In this paper, we extend the theoretical insights in \cite{HoYuanSunICC13}, as well as present new algorithmic solutions and results for utility maximization with data offloading. Our contributions are as follows.
Based on a unified framework for the problem of data offloading, we obtain fundamental properties on the computation, feasibility, and monotonicity of the load-coupling system.
For a given (small cell or WiFi) complementary network, we formulate a utility-maximization problem in which the users' demand can be served in either the (regular) cellular network or the complementary network, or concurrently in both networks.
We establish conditions for which the optimization problem has a feasible solution and is convex, and hence tractable to numerical computations.
We also propose a strategy to constrain the load to some maximum value, as required for practical implementation, and provide theoretical justification for the proposed algorithm.
Numerical results are obtained for both under-loaded and over-loaded networks, and can serve as a reference for the design of data offloading systems in practice.
The main tool we employ for analysis is based on the Perron-Frobenius theorem and other related results  \cite{Stanczak09}.

Section~\ref{sec:model} gives the system model of the load-coupled network.
Section~\ref{sec:exist} presents the fundamental properties of the load-coupled system.
Section~\ref{sec:doff} formulates the data offloading problem. Convexity analysis and an algorithm to constrain the maximum load are also given.
Numerical results are given in Section~\ref{sec:numerical}.
Section \ref{sec:con} concludes the paper.

{\it Notations}:
We denote a (tall) vector by a bold lower case letter, say $\bm{a}$. We denote a matrix by a bold capital letter, say $\bm{A}$, and denote its $(i,j)$th element by its lower case $a_{ij}$. 
We denote a {\em positive} matrix as $\bm{A}>0$ if $a_{ij}>0$ for all $i,j$.
Similarly, we denote a {\em non-negative} matrix as $\bm{A}\geq 0$ if $a_{ij}\geq 0$ for all $i,j$.
Similar definitions apply to vectors.


\section{System Model}\label{sec:model}

%
%


We consider a cellular network consisting of $n$ base stations that can interfere with each other. We focus on the downlink communication scenarios where base station $i\in \mathcal{N}\triangleq\{1,\cdots, n\}$ transmits with power $p_i\geq 0$.
We refer to cell $i$ interchangeably with base station $i$.
For notational convenience, we collect all power $\{p_i\}$ as vector
$\bm{p}>0$.

Each base station $i$ serves one unique group of users in set $\mathcal{J}_i$, where $|\mathcal{J}_i|\geq 1.$
User $j\in \mathcal{J}_i$ is served in cell $i$ up to a maximum rate of $D_{ij}$~nat.
Thus, the data can be interpreted as best-effort or elastic data to be served as much as possible subject to network conditions.
We also allow the users to be served in a {\em complementary cell}, to be introduced next. 


\subsection{Data Offloading to Complementary Network}\label{sec:complementary}

We shall consider data offloading, where the demand of every user can also be served in a complementary network. We assume a total of $n'$ complementary cells in the complementary network, denoted by the set $\mathcal{N}'=\{1,\cdots, n'\}$. Each complementary cell $i$ transmits with power $p_i'\geq 0$.

Specifically, we map every regular-cell user $j\in\mathcal{J}_i$ in the regular cell $i\in\mathcal{N}$ uniquely to a (virtual) complementary-cell user $b\in \mathcal{J}'_a$ in the complementary cell $a\in \mathcal{N}'$, via the mapping $(a,b)=\pi(i,j)$; note that both refers to the same physical user. 
We take the demands $d_{ij}$ and $d_{\pi(i,j)}'$ to be served in the regular and complementary cells, respectively,  as variables to be optimized, subject to the demand constraint
\be \label{eqn:dconstraint}
d_{ij}+d'_{\pi(i,j)}\leq D_{ij}, \;\; i\in\mathcal{N}, j\in\mathcal{J}_i.
\ee
The demand constraint ensures that the total demand served to each user is not more than the demand $D_{ij}$ requested. This is because any demand served beyond the requested amount may not benefit the users, yet consumes additional network resources at an increased cost for the cellular operator.
For notational convenience, we collect all demands $\{d_{ij}\}$ and $\{d_{\pi(i,j)}'\}$ as vectors
$\bm{d}\geq 0$ and $\bm{d}'\geq 0$, respectively.

We assume there is at least one user $j$ in cell $i$ with $d_{ij}>0$, otherwise  $p_i=0$ and so base station $i$ can be omitted; we make the same assumption for the complementary cells.
Thus without loss of generality, we have $p_i, p_i'>0$.




We consider two types of complementary network, consisting of either only small cells or WiFi cells.
For the case of small-cell offloading, both the regular cellular network and small-cell network use the same frequency band, hence the networks interfere with each other.
For the case of WiFi offloading, the frequency band used in the WiFi network is orthogonal to that of the cellular network, hence there is no mutual interference.
Our model can be easily generalized to the hybrid case consisting of a mixture of small cells and WiFi cells, with more cumbersome notations. For ease of exposure, we do not consider this hybrid case.

%
%
%
%
%
%

\subsection{Load Coupling Model}\label{sec:nocomplementary}

We first consider the load coupling model for the cellular network without any complementary network. The extension to the case with a complementary network is given in Section~\ref{sec:loadcouple_with_complem}.


Let $\loadv=[\load_1, \cdots, \load_n]$ be the load of the cellular network, where $0\leq \loadv\leq 1$. The  load $\load_i$, measures the fractional usage of resource in cell $i$. In LTE systems, the load can be interpreted as the expected fraction of the time-frequency resources that are scheduled to deliver data.
We model the SINR of user $j$ in cell $i$ as  \cite{SiFuFo09, MaKo10, SiominaYuan, FehskeFettweisICC12}
\be\label{eqn:sinr}
\sinr_{ij}(\loadv)= \frac{p_i g_{ij}}{\sum_{k\in \mathcal{N} \backslash \{i\}} p_k g_{kj} \load_k+\sigma^2 }
\ee
where $\sigma^2$ represents the noise power and $g_{ij}$ is the channel gain (or channel power) from base station $i$ to user $j$; note that $g_{kj}, k\neq i,$ here represents the channel gain from interfering base station $k$.
The SINR model \eqref{eqn:sinr} gives good approximation of more complicated cellular models \cite{FehskeFettweisICC12}.
Intuitively, $\load_k$ can be interpreted as the probability of receiving interference from cell $k$ on all the sub-carriers of the resource unit. Thus, the combined term $(p_k g_{kj} \load_k)$ is interpreted as the expected interference with expectation taken over time and frequency for all transmissions.

Since Gaussian-signalling is the worst-case noise distribution for mutual information \cite{Cover06}, an achievable rate is given by $r_{ij}=B \log(1+\sinr_{ij})$~nat/s per resource unit, where $B$ is the bandwidth for one resource unit and $\log$ is the natural logarithm. 
To deliver a demand of $d_{ij}$~nat for user $j$, the $i$th base station thus uses $x_{ij}\triangleq d_{ij}/r_{ij}$ resource units.
We assume that at total $M$ (time and frequency) resource units are available.
Summing the resource units over all users in cell $i$, we get the load for the cell as
\be
\load_i&=& \sum_{j\in\mathcal{J}_i} x_{ij}/M \\
&=&
\frac{1}{MB}
\sum_{j\in\mathcal{J}_i}\frac{d_{ij}}{\log\left(1+\sinr_{ij}(\loadv)\right)}
\triangleq {f}_i(\loadv)
\label{eqn:nonlinearprob0}
\ee
for $i\in \mathcal{N}$. 
For notational simplicity, we normalize $d_{ij}$ and $r_{ij}$ by the total amount of resource units $MB$.
Hence,  without loss of generality we let $MB=1$ in \eqref{eqn:nonlinearprob0}.

Let $\bm{f}(\loadv)=[f_1(\loadv), \cdots, f_n(\loadv)]^T$.
In vector form, we have
\be\label{eqn:nonlinearprob}
\loadv=\bm{f}(\loadv; \varspace)
\ee
where we have made the dependence of the load on the demand $\bm{d}$ and power $\bm{p}$ explicit.
We call \eqref{eqn:nonlinearprob} the {\em non-linear load coupling equation} (NLCE), as the load $\loadv$ appears in both sides of the equation and cannot be readily solved in closed-form.
To emphasize that a load is a solution of the NLCE, we denote the load as $\loadv^{\star}$ when necessary.
We say the load $\loadv^{\star}$ to be {\em feasible} if $\loadv^{\star}$ satisfies the NLCE {\em and} $\loadv^{\star} \geq 0$.
An algorithm that ensures that the load is also less than one (by reducing the demand) shall be considered in Section~\ref{sec:limitingload}.



\subsection{Load Coupling with Complementary Network}\label{sec:loadcouple_with_complem}

For the cellular network with a small-cell network, the two networks operate in the same frequency band and can be treated as one integrated network. 
Specifically, the set of $n$ base stations in the regular network is combined with the set of $n'$ base stations in the small-cell network to form a larger set of base stations of size $n+n'$.
All base stations can interfere with one another.

For the cellular network with a WiFi network, the two networks operate in different frequency bands.
We assume the WiFi network also submits to the load-coupling system relation.
That is, the NLCE holds for the cellular network as before, and also holds separately for the WiFi network by replacing $\loadv, \bm{d}, \bm{p}$ with the corresponding WiFi quantities denoted by  $\loadv', \bm{d}', \bm{p}'$.

We note that regardless of whether the complementary network is a small cell or WiFi network, the allocation of $\{d_{ij}, d_{\pi(ij)}'\}$ is coupled due to the constraint $d_{ij}+d_{\pi(ij)}'\leq D_{ij}$.

\section{Feasible Load: Fundamental Properties}\label{sec:exist}

We explore fundamental properties related to NLCE, namely, computation, existence and monotonicity of the load solution.
For clarity, we consider the regular cellular network without any complementary network; the results extend straightforwardly to the case with complementary network via the discussion in Section~\ref{sec:loadcouple_with_complem}.


\subsection{Computation}

Consider the following iterative algorithm. Starting from an arbitrary initial load $\loadv^0> 0$, define the $k$th iteration solution as
\be\label{eqn:algo}
\loadv^{k}=\bm{f}(\loadv^{k-1}; \varspace)
\ee
for $k=1,2,\cdots,K$, where $K$ is the total number of iterations. 
Lemma~\ref{lem:algo} ensures that $\loadv^{K}$ converges to the feasible load $\loadv^{\star}$ in the NLCE for large $K$.
The proof relies on the property of the standard interference function as defined in \cite{Yates95}.

\begin{lemma} \label{lem:algo}
Suppose a feasible load $\loadv^{\star}$ exists for the NLCE \eqref{eqn:nonlinearprob}. 
Then $\loadv^{K}$ converges to the unique fixed point solution $\loadv^{\star}$ as $K\rightarrow\infty$.
\end{lemma}
\begin{IEEEproof}
We sketch the proof given in  \cite{FehskeFettweisICC12}. After establishing that $\bm{f}(\cdot)$ is a standard interference function, Theorem~2 in \cite{Yates95} is applied to obtain the desired result.
\end{IEEEproof}

\begin{remark}[Asynchronous iteration]\label{rem:asyn_itn}
The iterative algorithm in \eqref{eqn:algo} is said to be {\em synchronous} \cite{Yates95} because all elements in vector $\loadv^k$  are obtained simultaneously. We may also consider the {\em asynchronous} version, in which a set of one or more elements are updated multiple times followed sequentially by other sets until all cells are updated at least once. By using Theorem~4 in \cite{Yates95}, we also obtain the convergence property in Lemma~\ref{lem:algo} with asynchronous iterations.
\end{remark}

The observation in Remark~\ref{rem:asyn_itn} is useful for implementation in practice, because the base stations can adapt their load in a distributed manner, and yet a feasible load can be obtained after sufficient number of iterations. Moreover, the asynchronous iteration will be used from an analytical viewpoint later, in the proof of Theorem~\ref{thm:d_inc_with_x}.

\subsection{Existence}

Before we compute the load as in Lemma~\ref{lem:algo}, we need to check if a feasible load exists.
Lemma~\ref{lem:equiv_existence} next states that feasibility can be checked by a simpler problem via a {\em linear} counterpart to the NLCE.

\begin{lemma}\label{lem:equiv_existence}
Given $\bm{d}$ and $\bm{p}$, a feasible load $\loadv^{\star}\geq 0$ exists for the NLCE \eqref{eqn:nonlinearprob} if and only if a solution $\loadv\geq 0$ exists in
\be\label{eqn:linearprob}
\loadv=\bm{H}(\varspace) \cdot \loadv + \bm{c}(\varspace).
\ee
Here, $\bm{c}(\varspace) \triangleq \bm{f}(\bm{0}_{n}; \varspace)$, where $\bm{0}_{n}$ is the length-$n$ all-zero vector, and $\bm{H}(\varspace) \geq 0$ is the real matrix with $(i,k)$th element
\be
{h}_{ik}=
\left\{
\begin{array}{ll}
0, & \mbox{if }i=k; \\
  (p_k  / p_i) \sum_{j\in\mathcal{J}_i} g_{kj} d_{ij}/ g_{ij}, & \mbox{if } i\neq k
\end{array}
\right.
\ee
for $1\leq i\leq n$ and $1\leq k\leq n$.
Note that $\bm{c}(\varspace)>0$ because at least one $d_{ij}$ in cell $i$ is positive.
\end{lemma}
\begin{IEEEproof}
From Theorem~8 and Theorem~11 in \cite{SiominaYuan}.
\end{IEEEproof}



%

Next, we treat $\bm{d}$ and $\bm{p}$ as variables to be optimized, so as to study how they affect the feasibility of the load.
Our main result is stated in Theorem~\ref{thm:1} below, which gives the necessary and sufficient condition for a feasible $\loadv$ to exist.

We make some preparation before stating the theorem. Let $\bm{\Lambda}(\bm{d})\geq 0$  be the $n$-by-$n$ real matrix with the $(i,k)$th element 
\be\label{eqn:def:lambda}
\lambda_{ik}=
\left \{
\begin{array}{ll}
0, & \mbox{if } i=k; \\
\sum_{j\in\mathcal{J}_i}  {g_{kj} d_{ij}}/{g_{ij}}, & \mbox{if } i\neq k
\end{array}
\right .
\ee
for $1\leq i \leq n$ and $1\leq k \leq n$.
We can therefore express the matrix $\bm{H}(\varspace)$ in \eqref{eqn:linearprob} as
\be\label{eqn:nicerep}
\bm{H}(\varspace) = \diag(\bm{p}) \cdot \bm{\Lambda}(\bm{d}) \cdot \diag(\bm{p})^{-1}
\ee
where $\diag(\bm{p})$ denotes the diagonal matrix with diagonal elements $\bm{p}$.
The effects of $\bm{p}$ and $\bm{d}$ are thus decoupled into three matrices, and so \eqref{eqn:linearprob} becomes
\be\label{eqn:linearprob2}
\widetilde{\loadv} =\bm{\Lambda}(\bm{d})  \widetilde{\loadv} + \bm{\widetilde{c}}(\bm{p},\bm{d})
\ee
where $\widetilde{\loadv}\triangleq\diag(\bm{p})^{-1} \loadv$ and $\bm{\widetilde{c}}(\varspace)\triangleq \diag(\bm{ p})^{-1} \bm{c}(\varspace).$

\begin{theorem}\label{thm:1}
Given $\bm{p}>0$ and $\bm{d}\geq 0$, a feasible load $\loadv^{\star}\geq 0$ for the NLCE \eqref{eqn:nonlinearprob} exists if and only if
\be\label{eqn:spectralradius}
r(\bm{\Lambda}(\bm{d})) < 1
\ee
where $r(\bm{\Lambda})$ is  the \emph{spectral radius} of matrix $\bm{\Lambda}$, defined as the absolute value of the largest eigenvalue of $\bm{\Lambda}$.
\end{theorem}
\begin{IEEEproof}
By Lemma~\ref{lem:equiv_existence}, it is sufficient to consider the linear counterpart \eqref{eqn:linearprob}, or equivalently \eqref{eqn:linearprob2}.
Since $\bm{p}>0$, every base station $i$ serves some positive demand and so $\sum_{j\in\mathcal{J}_i} d_{ij}>0$. Thus, $\bm{\Lambda}(\bm{d})\geq 0$ and 
$\bm{c}(\varspace)>0$.
Hence, applying the Perron-Frobenius theorem in \cite[Theorem~A.51]{Stanczak09} to \eqref{eqn:linearprob2}, we conclude that \eqref{eqn:spectralradius} is  necessary and sufficient for a feasible $\widetilde{\loadv}$ to exist in \eqref{eqn:linearprob2}.
%
\end{IEEEproof}

From \eqref{eqn:spectralradius}, the existence of a feasible load depends only on the demand vector $\bm{d}$, but not on the power $\bm{p}$.  This suggests the importance of data offloading by varying the demand, which is made explicit in Corollary~\ref{thm:importload}.

\begin{corollary}\label{thm:importload}
Suppose a feasible load does not exist for a given demand $\bm{d}\geq 0$ and power $\bm{p}>0$. Then no feasible load can exist by varying only $\bm{p}$. However, a feasible load always exist by varying $\bm{d}$.
\end{corollary}
\begin{IEEEproof}
The spectral radius $r(\bm{\Lambda}(\bm{d}))$ depends only on the demand $\bm{d}$. Hence, changing the power $\bm{p}$ does not affect the existence of the feasible load.
But scaling the demand vector uniformly by a positive factor allows the spectral radius to be scaled also by the same factor. Hence the spectral radius can always be made smaller than one by reducing the demand such that a feasible load exists.
\end{IEEEproof}



%

Motivated by Corollary~\ref{thm:importload}, subsequently we shall focus on the scenario where only the demand is varied, while the power is always taken to be fixed and positive. 

\subsection{Monotonicity of Load as a Function of Demand}

With power fixed, Theorem~\ref{thm:d_inc_with_x} shows that the load vector that satisfies the NLCE is a monotonic function of the demand vector.

\begin{theorem}\label{thm:d_inc_with_x}
Consider the NLCE \eqref{eqn:nonlinearprob} with power $\bm{p}$ fixed.
Given the demand vectors $\bm{d}'$ and $\bm{d}$ with $\bm{d}' \geq \bm{d}$ and $\bm{d}' \neq \bm{d}$, the corresponding NPCE load ${\loadv'}^{\star}$ and $\loadv^{\star}$ satisfy ${\loadv'}^{\star}>\loadv^{\star}$.
\end{theorem}
\begin{IEEEproof}
We sketch the proof; the details are given in Appendix~\ref{app:d_inc_with_x}. First, consider the case that only one element of $\bm{d}'$ is strictly greater than $\bm{d}$, with all other demand elements unchanged. Then we employ an asynchronous iteration in Remark~\ref{rem:asyn_itn} with initial load $\loadv^{\star}$. Upon convergence, we obtain ${\loadv'}^{\star}$ which can be shown to satisfy ${\loadv'}^{\star}>\loadv^{\star}$. Finally we consider the case where more than one element of $\bm{d}'$ is strictly greater than $\bm{d}$. In the proof, we apply the above argument successively to each element of the demand vector where the strict inequality holds.
\end{IEEEproof}

Theorem~\ref{thm:d_inc_with_x} also justifies our approach of focusing on a feasible load vector such that $\loadv^{\star}\geq 0$. Once feasibility is established, we can reduce the demand further to ensure that $0\leq \loadv^{\star}\leq 1.$

\section{Demand Offloading}\label{sec:doff}

%


We model the benefit of serving the demand in a network with offloading via three representative utility functions in Section~\ref{sec:utility}.
Next, we pose the optimization problem of maximizing the sum utility in Section~\ref{sec:optprob}, where the complementary network is either a WiFi or small-cell network. Then we investigate the convexity of the solution space, which affects the difficulty of numerical computations of the optimal solution, for $n=2$ base stations in Section~\ref{sec:n=2} and for arbitrary $n$ in Section~\ref{sec:arbi_users}. Finally, we propose an algorithm to limit the maximum optimal load to one in Section~\ref{sec:limitingload}.

As before, we shall consider feasible load such that $\loadv^{\star}\geq 0$ in this section. In Section~\ref{sec:limitingload}, we shall impose the additional constraint that the feasible load is less than one, i.e., $0\leq \loadv^{\star}\leq 1.$


\subsection{Utility for Maximization}\label{sec:utility}


Our objective is to maximize the sum utility
\be
\usum \triangleq \sum_{i\in\mathcal{N}} \sum_{j\in\mathcal{J}_i} k_{ij} U(d_{ij}) +  k_{\pi(i,j)}' U(d_{\pi(i,j)})
\ee
where $U(d)$ is the utility function for satisfying demand $d$. 
The weights $k_{ij}$ and $k_{\pi(i,j)}'$ take into account the combined priority of the user and the networks.
The utility function can be used to quantify the value of serving the demand $d$ to the cellular operator or user in terms of, for instance, the revenue collected from the access service, and
the fairness of serving the demand to multiple users within each cell type.
We note that the advantage of serving in either cell type can be quantified via the weights $k_{ij}$ and $k_{\pi(i,j)}'$.

To give insights, $U(d)$ is chosen to be the following representative functions, namely the linear ($\LIN$), logarithmic ($\LOG$), and double-logarithmic ($\DLOG$) utility functions:
\be
\LIN: \; & U(d) =& d,  \\
\LOG: \; & U(d) =& \log(d),  \\
\DLOG: \; & U(d) =&\log(\log(1+d)).
\ee
The utility functions are monotonically increasing and hence one-to-one functions.
The LIN utility models the scenario where serving an additional demand unit results in an additional unit of utility. For LOG utility, serving an additional demand unit of a user with a low demand results in more utility. Intuitively, this results in a fairer demand distribution among users but could result in a smaller revenue to the operator as less total demand is served. Thus, the LOG utility trades revenue maximization with user fairness. The DLOG utility further emphasize fairness, because it favours low-demand users even more.
We note that the last two utility functions would not assign zero demand to any user, because the sum utility is then negative infinity. The generalization to a broader class of functions is considered in Remark~\ref{rem:superlog} later.

For exposure, we make the {\em same-demand assumption} that every user $j$ in the same regular cell $i$ is served the same demand $d_{ij}=\widetilde{d}_i$.
Corresponding to the regular-cell user $j$ in cell $i$, we denote the complementary-cell user as $a(i,j)$ in complementary cell $b(i,j)$, i.e., $(a(i,j),b(i,j))=\pi(i,j)$.
For the complementary network, we also make the same-demand assumption, i.e., $d_{\pi(i,j)}'=\widetilde{d}_{a(i,j)}'$ for all $i,j$.
In effect, we focus on varying the cell-level demand vectors $\bm{\widetilde{d}}\triangleq [\widetilde{d}_1,\cdots, \widetilde{d}_n]^T$ and $\bm{\widetilde{d}}'\triangleq [\widetilde{d}_1',\cdots, \widetilde{d}'_{n'}]^T$.
From the demand constraint \eqref{eqn:dconstraint}, we get
\be\label{eqn:constraintdemand}
{d}_{ij}+{d}_{\pi(i,j)} = \widetilde{d}_{i}+\widetilde{d}_{a(i,j)} \leq D_{ij}, \forall j\in\mathcal{J}_i, i\in \mathcal{N}.
\ee
Since all cells are active with power vector  $\bm{p}>0$, we also have $\bm{\widetilde{d}}>0$ and $\bm{\widetilde{d}}'>0$.

\begin{remark}[Relaxing same-demand assumption]
For the case of $\LOG$ utility, the same-demand assumption can be slightly relaxed. Instead we assume more generally that each user $j\in \mathcal{J}_i$ in cell $i$ is allocated a demand of $d_{ij}=\alpha_{ij}\widetilde{d}_i$, where $\sum_{j\in\mathcal{J}_i} \alpha_{ij}=1$. Here,
$\alpha_{ij}$ is a fraction of the total demand $\widetilde{d}_i$ served in cell $i$. Thus, the user's achieved utility is $U(\alpha_{ij}\widetilde{d}_i)=\log(\widetilde{d}_i)+\log(\alpha_{ij})$.
With $\alpha_{ij}$'s fixed, it suffices to consider the first term $\log(\widetilde{d}_i)$ for the sum utility $\usum$. Hence the optimization problem is similar to the case under the same-demand assumption.
In general, however, the same-demand assumption is required for the subsequent convexity results to hold.
\end{remark}



\subsection{Optimization Problem}\label{sec:optprob}

\subsubsection{WiFi as Complementary Network}
We first formulate the optimization problem with WiFi as the complementary network.
Mathematically, our {\em data offloading problem} is
\be\label{p0}\IEEEyessubnumber
(P0)\;\;
\max_{\bm{\widetilde{d}}, \bm{\widetilde{d}}'}  &\;\; & \sum_{i\in\mathcal{N}} \sum_{j\in\mathcal{J}_i} k_{ij} U(\widetilde{d}_{i}) + k_{\pi(i,j)}' U(\widetilde{d}_{a(i,j)}')\\ \IEEEyessubnumber \label{p0:c1}
\mbox{\; s.t. } && \bm{\widetilde{d}}\in \mathcal{F}\triangleq \{\bm{\widetilde{d}}>0: r(\bm{\Lambda}(\bm{\widetilde{d}})) < 1\} \\ \IEEEyessubnumber \label{p0:c2}
&& \bm{\widetilde{d}}'\in \mathcal{F}'\triangleq \{\bm{\widetilde{d}}'>0: r(\bm{\Lambda}'(\bm{\widetilde{d}}')) < 1\} \\
\IEEEyessubnumber \label{p0:c3}
 &&    \widetilde{d}_i+\widetilde{d}_{a(i,j)}' \leq D_{ij}, \; \forall j\in\mathcal{J}_i, i \in \mathcal{N}
\ee
where $\bm{\Lambda}, \bm{\Lambda}'$ correspond to \eqref{eqn:def:lambda} for the regular cellular network and the WiFi network, respectively.
The constraints \eqref{p0:c1}, \eqref{p0:c2} follow from Theorem~\ref{thm:1} and the discussion for WiFi network in Section~\ref{sec:loadcouple_with_complem}; we call $\mathcal{F}$ and $\mathcal{F}'$ the {\em feasibility sets}.
The last constraint is due to \eqref{eqn:constraintdemand}.
We note that a solution always exists because we can always reduce the demand to arbitrarily close to zero, so as to satisfy constraints \eqref{p0:c1} and \eqref{p0:c2} (see the proof of Corollary~\ref{thm:importload}) and also to satisfy constraint \eqref{p0:c3}.

For convenience, we transform $\widetilde{d}_i$ to $y_i=U(\widetilde{d}_i)$ for $i\in\mathcal{N}$ and let $\bm{\widetilde{d}} = [U^{-1}(y_1), \cdots, U^{-1}(y_n)]^T \triangleq \bm{g}(\bm{y}).$ The inverse $U^{-1}(\cdot)$ always exists because $U(\cdot)$ is a monotonic function.
Similarly for the WiFi cells, let $y'_i=U(\widetilde{d}'_i)$ for $i\in\mathcal{N}$ and $\bm{\widetilde{d}}' = [U^{-1}(y'_1), \cdots, U^{-1}(y'_n)]^T \triangleq \bm{g}(\bm{y}').$
Let $k_i\triangleq \sum_{j\in\mathcal{J}_i} k_{ij}$.
We make similar definitions for $y_i'$ and $k_i'$ corresponding to the complementary cells.
Our {\em transformed data offloading} problem is then
\be\label{p1}\IEEEyessubnumber
(P1)\;\;
\max_{\bm{y},\bm{y}'}  &\;\; & \sum_{i\in\mathcal{N}}  k_i y_i + k_i' y_i'
\\ \IEEEyessubnumber \label{p1:c1}
\mbox{\; s.t. } && \bm{y}\in \mathcal{\widetilde{F}}\triangleq \{ \bm{y}\in \mathcal{Y}^n: r(\bm{\Lambda}(\bm{g}(\bm{y}))) < 1\}
\\ \IEEEyessubnumber \label{p1:c1a}
                && \bm{y}'\in \mathcal{\widetilde{F}}'\triangleq \{ \bm{y}'\in \mathcal{Y}^{n'}: r(\bm{\Lambda}'(\bm{g}(\bm{y}'))) < 1\}
\\
 &&  U^{-1}(y_i) + U^{-1}(y_{a(i,j)}')\leq D_{ij}, \; \forall j\in\mathcal{J}_i, i \in \mathcal{N}
\nonumber \\
\IEEEyessubnumber \label{p1:c2}
\ee
where $\mathcal{Y}^n$ of dimension $n$ is defined as the set of positive vectors for $\LIN$ utility, and as the set of real vectors for $\LOG$ and $\DLOG$ utilities.
Here $\mathcal{\widetilde{F}}$ (and similarly $\mathcal{\widetilde{F}}'$) is the transformed feasibility set with complement set denoted as $\mathcal{\widetilde{F}}^c = \mathcal{Y}^n \setminus \mathcal{\widetilde{F}}.$
For $\LIN$ utility, Problem $P1$ is the same as Problem $P0$, and thus $\mathcal{\widetilde{F}}=\mathcal{F}$.

We note that the objective function is always linear in $\bm{y}$ and $\bm{y}'$.
For any of the three utility functions, it can be checked that the set of $\bm{\widetilde{y}}, \bm{\widetilde{y}}'$ subject only to \eqref{p1:c2} is convex. Now if $\mathcal{\widetilde{F}}$ (and similarly $\mathcal{\widetilde{F}}'$) is a convex set, $P1$ is a convex optimization problem for which numerically efficient solvers exist \cite{boyd}. In summary, it is sufficient to obtain conditions for which $\mathcal{\widetilde{F}}$  is convex, in order to ascertain if problem $P1$ is convex.

To account for individual demand constraints imposed by each network, we may also impose, in Problem~$P0$, additional constraints on $\widetilde{d}_i$ and $\widetilde{d}_{a(i,j)}'$ such that the variables do not exceed some fixed constants. It can be easily checked that such constraints do not affect the convexity of the solution space in Problem~$P1$.

\subsubsection{Small cell as Complementary Network}
The optimization problem is similar to problem $P0$, except that the feasibility sets in \eqref{p0:c1} and \eqref{p0:c2} are merged into a single feasibility set subject to $r(\bm{\Lambda}''(\bm{\widetilde{d}}, \bm{\widetilde{d}}')) < 1$ where $\bm{\Lambda}''$ includes the base stations of both the regular cells and the small cells, see Section~\ref{sec:loadcouple_with_complem}.
By similar arguments as before, for the transformed data offloading problem to be convex, it suffices to check if $\mathcal{\widetilde{F}}$ that corresponds to $\bm{\Lambda}''(\bm{\widetilde{d}}, \bm{\widetilde{d}}')$ is convex.

\subsection{Two Base Stations}\label{sec:n=2}

To gain some understanding for the convexity of $\mathcal{\widetilde{F}}$, let us study the case of $n=2$ base stations.
It can be verified that if we write $\bm{\Lambda}(\bm{d})=\left[\begin{array}{cc}
0 & \beta\\
\beta' & 0
\end{array} \right]$, then the unit eigenvectors and corresponding eigenvalues of $\bm{\Lambda}(\bm{d})$ are
$
\left\{ \xi[\sqrt{\beta}, \sqrt{\beta'}]^T, \sqrt{\beta\beta'}\right\}, \left\{ [-\sqrt{\beta}, \sqrt{\beta'}]^T, -\sqrt{\beta\beta'}\right\}
$
with $\xi\triangleq (\beta+\beta')^{-1/2}$.
The spectral radius can then be obtained in closed-form as $r(\bm{\Lambda}(\bm{d})) = \sqrt{\beta\beta'}$.
Thus the (non-transformed) feasibility set $\mathcal{F}$ 
is given by all $\bm{\widetilde{d}}=[\widetilde{d}_1, \widetilde{d}_2]^T$ that satisfies
\be\label{eqn:cor:n2}
\widetilde{d}_1 \widetilde{d}_2 \left(\sum_{j\in\mathcal{J}_1}  \frac{g_{2j}}{g_{1j}}\right) \left(\sum_{j\in\mathcal{J}_2}  \frac{g_{1j}}{g_{2j}} \right)< 1. 
\ee
Clearly, the feasibility set depends on the channel gains in a non-linear manner.
We note that the optimal $(\widetilde{d}_1,\widetilde{d}_2)$ lies on the inner boundary of $\mathcal{\widetilde{F}}$, since to maximize the objective function we must choose $\widetilde{d}_1$ or $\widetilde{d}_2$, or both, to be as large as possible. Moreover, the following observations can be made for the three utility functions.


\newcommand{\Scale}{0.55}
\newcommand{\SPACING}{-1.75cm}

\begin{figure}
\centering
\hspace{\SPACING}
\subfigure[][$\LIN$ utility.]
{\includegraphics [scale=\Scale] {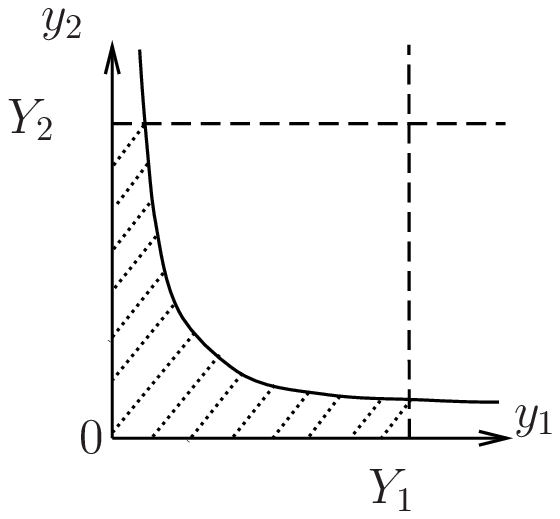}\label{fig:linearobj}}
\hspace{\SPACING}
\subfigure[][$\LOG$ utility.]
{\includegraphics [scale=\Scale] {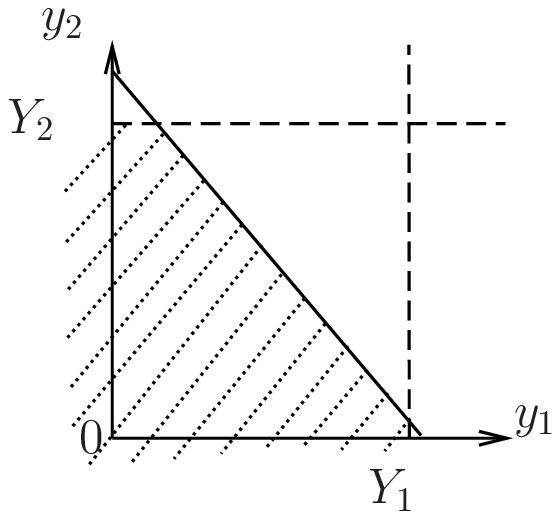}\label{fig:logobj}}
\hspace{\SPACING}
\subfigure[][$\DLOG$ utility.]
{\includegraphics [scale=\Scale] {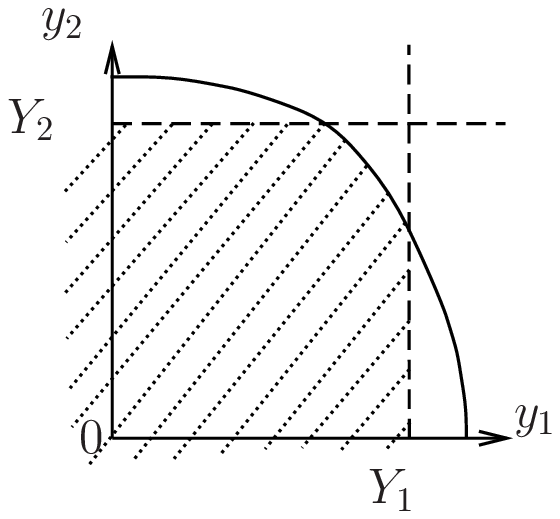}\label{fig:superlogobj}}
\hspace{\SPACING}
\caption{Transformed feasibility set (shaded) for different utility objective functions. After transformation, the objective function is always linear.} \label{fig:allobj}
\end{figure}

\underline{$\LIN$ utility:} The transformed feasibility set $\mathcal{\widetilde{F}}=\mathcal{F}$ is unchanged; see Fig.~\ref{fig:linearobj}. We include constraint \eqref{p1:c2} which can be written as $y_i\leq Y_i, i=1,2$; the actual value for $Y_i$ depends on the optimal demand vector for the complementary network. To maximize the sum utility, clearly the optimal solution is to assign either $y^{\star}_1=Y_1$ or $y^{\star}_2=Y_2$, i.e., an extreme solution. Moreover, the optimal solution is unique.

\underline{$\LOG$ utility:} 
The transformed feasibility set $\mathcal{\widetilde{F}}$, including the constraint \eqref{p1:c2}, is a polytope; see Fig.~\ref{fig:logobj}. To maximize the objective function $k_1 y_1 + k_2 y_2$, an optimal solution is given by the boundary extreme solution. This conclusion is similar to the linear utility case, except that the optimal solution is unique only if $k_1\neq k_2$. 

\underline{$\DLOG$ utility:} 
The transformed feasibility set $\mathcal{\widetilde{F}}$, including the constraint \eqref{p1:c2}, is strictly convex; see Fig.~\ref{fig:superlogobj}. To maximize the objective fuction $k_1 y_1 + k_2 y_2$, the optimal solution is not necessarily an extreme solution, but is always unique. This suggests the fairest data offloading, as neither of the demands is likely to be very small.

In the next section, we shall use more sophisticated analytical tools to shed further insight on the convexity of the transformed feasibility set $\mathcal{\widetilde{F}}$ for any $n$.


\subsection{Arbitrary Number of Base Stations}\label{sec:arbi_users}

For larger $n$, the spectral radius cannot be computed in closed-form, and it is expected that the dependence on the channel gains remains non-linear and complicated.
Nevertheless, an efficient numerical approach is warranted for arbitrary number of base stations $n$.
Theorem~\ref{thm:general} states the convexity of the feasibility set $\mathcal{\widetilde{F}}$ or its complement $\mathcal{\widetilde{F}}^c$.

\begin{theorem}\label{thm:general}
The following convexity results hold.

\underline{$\LIN$ utility:}  $\mathcal{\widetilde{F}}^c$ is convex for $n=2$. But $\mathcal{\widetilde{F}}^c$ is generally not convex for $n\geq 3$.

\underline{$\LOG$ utility:}  $\mathcal{\widetilde{F}}$ is convex for $n=2$. Moreover, $\mathcal{\widetilde{F}}$ is strictly convex for $n\geq 3$.

\underline{$\DLOG$ utility:}  $\mathcal{\widetilde{F}}$ is strictly convex for any $n\geq 2$. 

\end{theorem}
\begin{IEEEproof}
The proof for $n=2$ for all cases was given in Section~\ref{sec:n=2}. We now consider the case $n\geq 3$ by applying the results in \cite{Stanczak09}, which are closely related to the well-known Perron-Frobenius theorem.
First, note that we can express
\be\label{eqn:tildeLambdalinear}
\bm{\Lambda}(\bm{g}(\bm{y}))
= \diag(g(y_1), \cdots, g(y_n)) \; \widetilde{\bm{\Lambda}}
\ee
where $g(y_i)$ is the $i$th element of $\bm{g}(\bm{y})$ and the $(i,k)$th element of $\widetilde{\bm{\Lambda}}$ is
\be\label{eqn:def:lambdatilde}
\widetilde{\lambda}_{ik}=
\left \{
\begin{array}{ll}
0, & \mbox{if } i=k; \\
\sum_{j\in\mathcal{J}_i}  {g_{kj} }/{g_{ij}}, & \mbox{if } i\neq k
\end{array}.
\right .
\ee

\underline{$\LIN$ utility:} We have $g(y)=y$.
Applying \cite[Theorem~1.60]{Stanczak09} known as the linear mapping case to \eqref{eqn:tildeLambdalinear}, we obtain that $\mathcal{\widetilde{F}}^c$ is in general not convex.

\underline{$\LOG$ utility:} We have $g(y)=\exp(y)$.
The matrix structure in \eqref{eqn:tildeLambdalinear} is referred to as the exponential mapping case in \cite{Stanczak09}. Moreover, $\widetilde{\bm{\Lambda}}$ and $\widetilde{\bm{\Lambda}}\widetilde{\bm{\Lambda}}^T$ are irreducible; see Lemma~\ref{lem:irreducible} 
with definition of irreducibility in the Appendix~\ref{app:irreducible}. These two conditions allow us to apply \cite[Theorem~1.63]{Stanczak09} to show that $\mathcal{\widetilde{F}}$ is strictly convex for $n\geq 3$

\underline{$\DLOG$ utility:} We have $g(y)=\exp(\exp(y))-1$.
The following inequality holds after some calculus and algebraic manipulations:
\be\nonumber
 &&d  \frac{\partial^2 U(d)}{\partial d^2} + \frac{\partial U(d)}{\partial d} \nonumber \\
 &=& \frac{\partial U(d)}{\partial d} \left(1-\frac{d}{1+d}\left(1+\frac{1}{\log(1+d)}\right)\right) \nonumber \\
&< & \frac{\partial U(d)}{\partial d}\left(1-\frac{d}{1+d}\left(1+\frac{1}{d}\right) \right) = 0
\label{eqn:gen_doublelog}
\ee
where the above inequality is due to $\log(1+d)< d$ for $d>0$.
From Lemma~\ref{lem:logconvex} in Appendix~\ref{app:logconvex} with $x$ and $f(x)$ replaced by $d$ and $U(d)$, respectively,
the inverse of $U(d)$, i.e., $g(y)$, is strictly log-convex.
Since all diagonal elements of $\diag(\bm{g}(\bm{y}))$ are strictly log-convex, by \cite[Corollary 1.46]{Stanczak09}, it follows that $\mathcal{\widetilde{F}}$ is strictly convex.
%
\end{IEEEproof}


%
%

The number of users does not significantly affect the complexity of the optimization problem $P1$, due to the same-demand assumption that we have imposed. Instead, the complexity of the optimization problem depends on $n$, the number of transmitters, e.g., base stations or access points. Assuming LOG or DLOG utility is used, Theorem~\ref{thm:general} states that the feasibility set is convex, and hence the complexity for large $n$ is still manageable with the use of convex optimization techniques \cite{boyd}.

\begin{remark}[Generalizing Utility Function]\label{rem:superlog}
Theorem~\ref{thm:general} applies to a more general class of utility function $U(d)$. Specifically, if the utility function satisfies $ d \frac{\partial^2 U(d)}{\partial d^2} + \frac{\partial U(d)}{\partial d} <0$ for $n\geq 3$, then $\mathcal{\widetilde{F}}$ is strictly convex. We note that $\DLOG$  is a special case, see \eqref{eqn:gen_doublelog}.
This conclusion follows immediately from the proof for Theorem~\ref{thm:general}, in which Lemma~\ref{lem:logconvex} was used to show that $g(y)$ is strictly log-convex. 
Moreover, it follows that $g(y)$ is convex and so the constraint \eqref{p1:c2} is convex as its left-hand side is a sum of convex functions.
Hence the optimization problem $P1$ is a convex optimization for this general class of utility functions.
\end{remark}

\subsection{Algorithm to Limit Maximum Load}\label{sec:limitingload}

So far in our analysis, we consider feasible load $\loadv^{\star}\geq 0$, which holds if the spectral-radius constraint is strictly less than one. In practice, the load cannot exceed one, due to limited availability of network resources. 
To impose a constraint $0\leq \loadv^{\star}\leq 1$ explicitly in Problem~$P0$ however appears challenging.
Instead, in this section, we propose an iterative algorithm that reduces the demand such that $0\leq \loadv^{\star}\leq 1$.

\subsubsection{Preliminaries}

Let us consider the following optimization problem that is generalized from Problem~$P0$.
Define Problem~$Q(\rho)$, where $0\leq \rmaxone\leq 1$ is an optimizing variable, to be the same as Problem~$P0$ but with the spectral-radius constraints \eqref{p0:c1} and \eqref{p0:c2} replaced by $r(\bm{\Lambda}(\bm{g}(\bm{y}))) < \rmaxone$ and $r(\bm{\Lambda}'(\bm{g}(\bm{y}'))) < \rmaxone$, respectively.
We denote the corresponding feasibility sets as $\mathcal{F}(\rho)$ and $\mathcal{F}'(\rho)$, respectively.
Clearly, Problem~$Q(\rmaxone)$ specializes to Problem~$P0$ if $\rmaxone=1$.
Corresponding to Problem~$Q(\rmaxone)$, the optimal demand vector and load vector are denoted respectively as $\bm{d}^{\star}(\rmaxone)$ and $\loadv^{\star}(\rmaxone)$ for the regular cellular network, and similarly ${\bm{d}'}^{\star}(\rmaxone)$ and ${\loadv'}^{\star}(\rmaxone)$ for the WiFi network.
Finally, we denote the maximum optimal load as $\lmax(\rmaxone)\triangleq \max\{ \load^{\star}_i(\rmaxone),  {\load'}^{\star}_j(\rmaxone), i\in \mathcal{N}, j\in \mathcal{N'} \}$. Thus, $\loadv^{\star}(\rho) \leq 1$ if and only if $\lmax(\rho)\leq 1$.

We note that all the analysis so far for Problem~$P0$ apply also for Problem~$Q(\rho)$, independent of the actual value of $\rho$.
Thus, the numerical solution for Problem~$Q(\rho)$ can be obtained similarly as for Problem~$P0$.
It is useful to note that $\loadv^{\star}(\rho)$ and $\lmax(\rho)$ with $\rho=1$ corresponds to the optimal values for the special case of Problem~$P0$.


If $\lmax(1)\leq 1$, then $\loadv^{\star}(1)$ is an optimal solution for Problem~$P0$ \emph{and} satisfies the required constraint $0\leq \loadv\leq 1$.
Henceforth, we assume that $\lmax(1)>1$. In the following, we first propose an algorithm such that the final load vector satisfies $0\leq \loadv\leq 1$, followed by the theoretical justifications.

\subsubsection{Algorithm}

To ensure the load is limited by one, we propose to use the demand vector $\bm{d}^{\star}(\rho)$ corresponding to the load vector solution $\loadv^{\star}(\rho)$ in Problem~$Q(\rho)$, where $\rho$ is determined by the solution of the following optimization problem:
\be\label{p2}
(P2)\;\;
\max_{0\leq \rho<1} \rho  \\
\mbox{\; s.t. } && \lmax(\rho) \leq 1.
\ee
That is, $\rho$ is the largest possible value such that $0\leq \loadv^{\star}(\rho)\leq 1$.
In general, $\lmax(\rho)$ is not a monotonic function of $\rho$. For example, see Fig.~\ref{fig:increasing_demand}, where the detailed scenario setup is described in Section~\ref{sec:numerical}. Nevertheless, since we have reduced the optimization to only one variable, an exhaustive search based on a finely-quantized interval over $0\leq \rho<1$ can be performed to solve Problem~$P2$, where for each $\rho$ Problem $Q(\rho)$ is solved. This method shall be employed to obtain numerical results in  Section~\ref{sec:numerical}.

\subsubsection{Theoretical Basis}\label{sec:theo_basis}

The theoretical basis for the above algorithm stems from Theorem~\ref{thm:U_inc} and Theorem~\ref{thm:r} below.
Theorem~\ref{thm:U_inc} ensures that the highest possible sum utility is achieved for Problem~$Q(\rho)$ if we choose $\rho$ to be as large as possible.
Theorem~\ref{thm:r} ensures the existence of a solution in Problem~$P2$  under the equal-demand assumption.

\begin{theorem}\label{thm:U_inc}
Denote the optimal sum utility value for Problem~$Q(\rho)$ as $\usum(\rho), 0\leq \rho\leq 1$. Then $\usum(\rho)$ is a strictly increasing function of $\rho$.
\end{theorem}
\begin{IEEEproof}
Let $\widetilde{\bm{d}}'=\widetilde{\bm{d}}+\bm{e}, \bm{e}\geq 0$. It can be easily checked from definition \eqref{eqn:def:lambda} that $\bm{E}\triangleq \bm{\Lambda}(\widetilde{\bm{d}}')-\bm{\Lambda}(\widetilde{\bm{d}}) \geq 0$, with equality if and only if $\bm{e}=0$. Similar to the proof that $\widetilde{\bm{\Lambda}}$ is irreducible in Lemma~\ref{lem:irreducible}, it can be shown that $\bm{\Lambda}(\widetilde{\bm{d}})\geq 0$ is irreducible. Thus, we can apply Lemma~\ref{lem:r_inequality} in Appendix~\ref{app:r_inequality} to get $r(\bm{\Lambda}(\widetilde{\bm{d}}'))=r(\bm{\Lambda}(\widetilde{\bm{d}})+\bm{E})\geq r(\bm{\Lambda}(\widetilde{\bm{d}}))$, with equality if and only if $\bm{e}=0$.
Thus, $\widetilde{\bm{d}}'\geq \widetilde{\bm{d}}$ if and only if $r(\bm{\Lambda}(\widetilde{\bm{d}}'))\geq r(\bm{\Lambda}(\widetilde{\bm{d}}))$.
This implies that the feasibility set $\mathcal{F}(\rho)$ (and similarly for $\mathcal{F}'(\rho)$) satisfies
$\mathcal{F}(\rho_1)\subset \mathcal{F}(\rho_2)$ for $\rho_1< \rho_2$.
Thus, $\usum(\rho_1)< \usum(\rho_2)$ for $\rho_1< \rho_2$, i.e.,  $\usum(\rho)$ is a strictly increasing function.
\end{IEEEproof}

For illustration, we plot the sum utility $\usum(\rho)$ (added by a constant such that it becomes positive) as a function of $\rho$ in Fig.~\ref{fig:increasing_demand}.
In accordance with Theorem~\ref{thm:U_inc}, the sum utility is increasing with $\rho$.



\renewcommand{\Scale}{0.85}

\begin{figure}
\centering
\includegraphics [scale=\Scale] {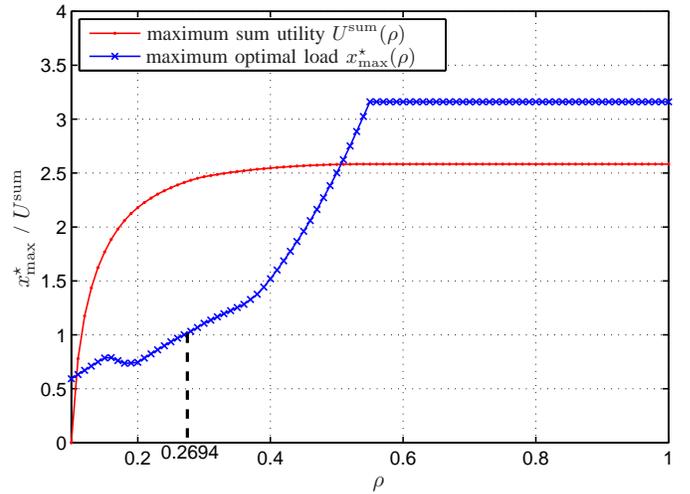}
\caption{The graph of maximum optimal load $\lmax$ and optimal objective value $\usum$ over $\rho$. The value $\rho^{\star}=0.2694$ corresponds to $\lmax(\rho^{\star})=1$.} \label{fig:increasing_demand}
\end{figure}


\begin{theorem}\label{thm:r}
Consider Problem~$Q(\rho)$ where $\lmax(\rmaxone) >1$ for $\rmaxone=1$. Then there exists an optimal load vector $\loadv^{\star}(\rho)$ such that $\lmax(\rho)= 1$ for some $0<\rmaxone<1$.
\end{theorem}
\begin{IEEEproof}
%
From the proof of Theorem~\ref{thm:U_inc}, the feasible set $\mathcal{\widetilde{F}}(\rmaxone)$ becomes strictly smaller as $\rho$ decreases. From Theorem~\ref{thm:d_inc_with_x}, the load vector is a monotonic function of the demand vector. Thus, every load vector corresponding to a demand vector in the feasible set also decreases in value, as $\rho$ decreases.
For sufficiently small $\rmaxone\rightarrow 0$, all the elements of the optimal demand must approach the all-zero vector and thus $\lmax\rightarrow 0$. By continuity, there exists $\lmax=1$ for some $0<\rmaxone<1$.
\end{IEEEproof}

For illustration, we plot the maximum optimal load $\loadv^{\star}(\rho)$ as a function of $\rho$ in Fig.~\ref{fig:increasing_demand}. We note that, in contrast to $\usum(\rho)$, $\loadv^{\star}(\rho)$ is not necessarily an increasing function of $\rho$. Nevertheless, there exist $\lmax(\rho)= 1$ as $\rho$ is decreased from $\rho=1$, in accordance with Theorem~\ref{thm:r}.
From Fig.~\ref{fig:increasing_demand}, we see that the largest $0\leq \rho<1$ such that $\lmax(\rho)=1$ is given by $\rho=\rho^{\star}=0.2694$.
Thus, this gives the solution for Problem~$P2$. The corresponding demand and load allocation is shown as Fig.~\ref{fig:demand_d05_LOG} later in Section~\ref{sec:numerical}.

\section{Numerical Results}
\label{sec:numerical}

\renewcommand{\Scale}{0.65}

\begin{figure}
\centering
\includegraphics [scale=\Scale] {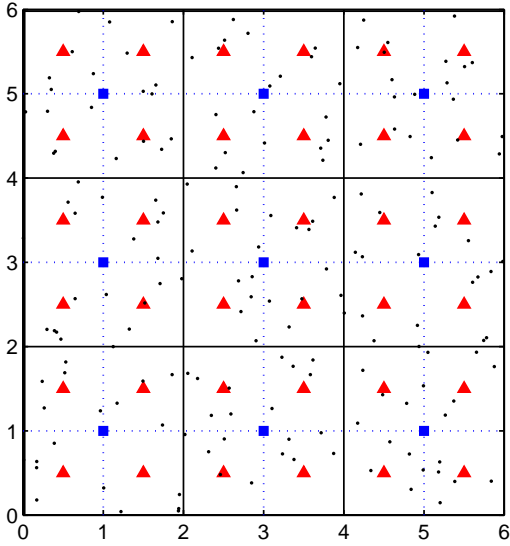}
\caption{Network configuration: base stations, access points and users are shown with the blue squares, red triangles, and black dots, respectively.} \label{fig:network}
\end{figure}

In this section, unless otherwise specified, we obtain numerical results assuming the utility function is the $\LOG$ utility. The optimization problem $P1$, and more generally $Q(\rho)$, is convex. This is because the objective function is linear and the constraint set is convex due to Theorem~\ref{thm:general} for the case of $\LOG$ utility. Thus, the optimal demand vectors $\widetilde{\bm{d}}^{\star}, \widetilde{\bm{d}}^{\hspace{0.075cm}'\hspace{-0.075cm}\star}$ can be solved efficiently by standard numerical solvers. Specifically, we use the active-set algorithm with the $\mathsf{fmincon}$ function in the \textsc{Matlab} software.
The optimal load vectors ${\loadv}^{\star}$ and ${\loadv'}^{\star}$ are then computed using a synchronous or asynchronous iterative algorithm according to Lemma~\ref{lem:algo} and Remark~\ref{rem:asyn_itn}, respectively.

Our theoretical result and numerical approach apply regardless of where the base stations, access points and users are deployed. For ease of viewing the numerical results, we  position the base stations and access points at equal distance apart, while the users are at arbitrarily but fixed locations (obtained by the realizations from a uniform distribution).
For the network configuration, we assume all cells are square in shape. The regular cellular network consists of $n=9$ cells, where each square cell is of two unit length. The cells are arranged uniformly as shown in Fig.~\ref{fig:network}. A base station is placed in the centre of each regular cell, shown as a blue square in Fig.~\ref{fig:network}. In each regular cell, there are four disjoint square WiFi cells each of unit length, making a total of $n'=36$ WiFi cells. Within each WiFi cell, there are $5$ users. A WiFi access point is placed in the center of each WiFi cell, shown as a red triangle. Every user is served by the WiFi cell and the base station cell that it resides in. Thus, every access point can support up to $5$ users while every base station can support up to $20$ users.
We make the same-demand assumption that all users in the same (regular or WiFi) cell are allocated the same demand.

We use the same weight $k_{ij}=1$ for all base stations, and the weight $k_{ab}'=1/4$ for all access points; the difference in weights is used to account for the fact that the number of access points is four times the number of base stations.
We set the (normalized) transmission power of every regular cell as $100$, the transmission power of every WiFi cell as $1$, and the noise variance as $0.01$.
The channel gain from the $i$th regular cell to the $j$th user is fixed as $g_{ij}=z_{ij}^{-\kappa}$ where $z_{ij}$ is the distance between transmitter $i$ and receiver $j$, and $\kappa=4$ is the path loss exponent. The channel gains for the WiFi cells are obtained similarly. 

\renewcommand{\Scale}{0.6}
\begin{figure}
\centering
\subfigure[][Optimal load allocation.]
{\includegraphics [scale=\Scale] {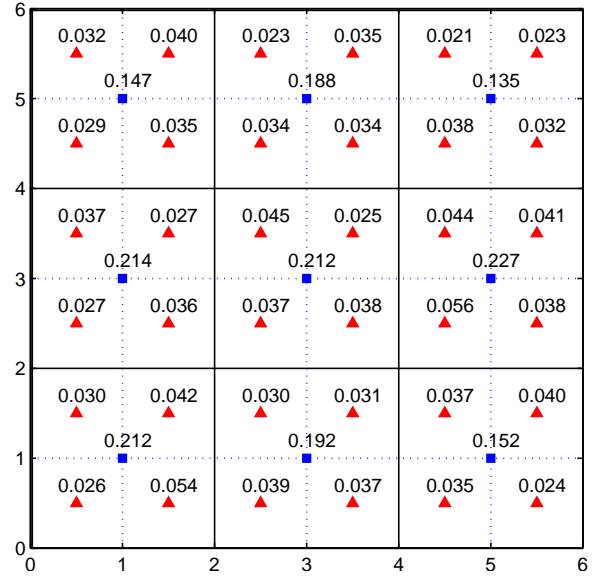}
\label{fig:demand_d01_load}}
\;\;
\subfigure[][Optimal demand allocation.]
{\includegraphics [scale=\Scale] {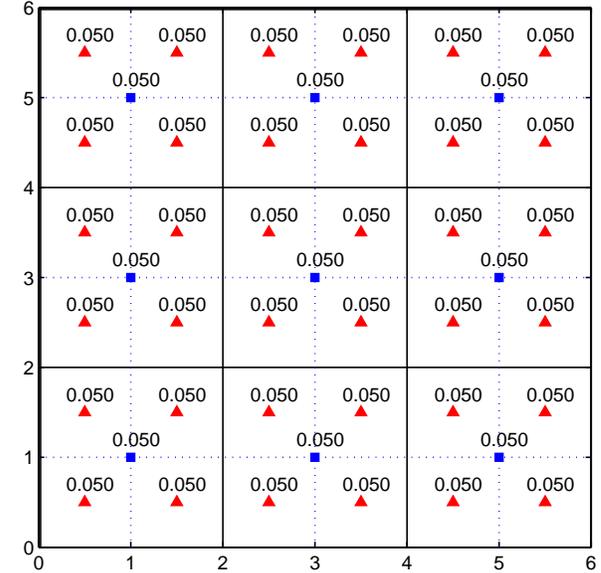}
\label{fig:demand_d01_demand}}
\caption{Optimal allocation with LOG utility and maximum demand fixed as $0.1$. The spectral radius is subject to a maximum constraint of $1$; the maximum load value turns out to be less than one.} \label{fig:demand_d01}
\end{figure}

\subsection{Low Maximum Demand}
In our first numerical experiment, we fix the maximum demand to be $D_i=0.1, i\in\mathcal{N}$.
Solving for Problem~$P1$ numerically, we obtain the optimal demand and load allocations as indicated in Fig.~\ref{fig:demand_d01} besides the positions of the base stations and access points.
From Fig.~\ref{fig:demand_d01_load}, all cells are operating below full load, i.e., $\lmax\leq 1$.
We observe that the load allocation in Fig.~\ref{fig:demand_d01_load} is non-uniform, due to the non-uniform user distribution as shown in Fig.~\ref{fig:network}.
From Fig.~\ref{fig:demand_d01_demand}, the optimal demand to be served by every regular cell and WiFi cell is the same, given by $d^{\star}=0.05$~nat.
Thus, all users are served the maximum rate of $D_i=0.1$~nat in total, with $d^{\star}$ contributed by the regular cell and another $d^{\star}$ contributed equally by the WiFi cell.
The reason for such a uniform distribution of the optimal demand follows.
We observe that the optimal demand is also given by $d^{\star}=0.05$ if we maximize the sum utility without the spectral-radius constraints \eqref{p1:c1} and \eqref{p1:c1a} (not shown here). This implies that the spectral-radius constraints are in fact not active in the original problem, i.e., the demands can be treated as unconstrained variables without loss of optimality.
For our choice of $k_{ij}=1,k_{ab}'=1/4$ and with one base station for every four access points, the optimal demand is thus uniform for all the access points and the base stations.
In further numerical experiments where the weights $k_{ij}, k_{ab}'$ are changed (not shown here), we observe that the optimal demand is not necessarily uniform, i.e., the values are different for the base stations and the access points.
Nevertheless, we make the consistent observation that the same optimal demand is obtained whether with or without the spectral-radius constraints; this is consistent with the earlier observation that the maximum load has not exceeded one.


\subsection{High Maximum Demand}
Next, we increase the maximum demand to $D_i = 0.45, i\in\mathcal{N}$. With this high maximum demand, we shall see that the spectral-radius constraint becomes active, and the maximum demand requested by the users cannot be achieved.

Solving for Problem~$P1$ numerically, we obtain the maximum optimal load as $\lmax(\rho)>1$ with $\rho=1$.
Thus, some of the cells are overloaded and the optimal demand vector $\bm{d}^{\star}(\rho)$ with $\rho=1$ cannot be practically implemented. To reduce the load, we solve Problem~$P2$ via the algorithm proposed in Section~\ref{sec:limitingload}. In this algorithm, we obtain $\rho^{\star}$ given by the largest $\rho$ in Problem~$Q(\rho)$ such that the corresponding maximum optimal load $\lmax(\rho)=1-\epsilon$ where $\epsilon>0$ is close to zero.
The sum utility $\usum$ and the maximum optimal load $\lmax$ are plotted as functions of $\rho$ in Fig.~\ref{fig:increasing_demand}. For ease of viewing, $\usum$ has been increased by a constant value.
From Fig.~\ref{fig:increasing_demand}, the optimal $\rho$ is $\rho^{\star}=0.2694$. 
The average per-user demand achieved, obtained by averaging the sum demand over all users, is then given by $0.4449$~nat, which is less than the maximum demand of $D_i = 0.45$.

\begin{figure}
\centering
\subfigure[][Optimal load allocation.]
{\includegraphics [scale=\Scale] {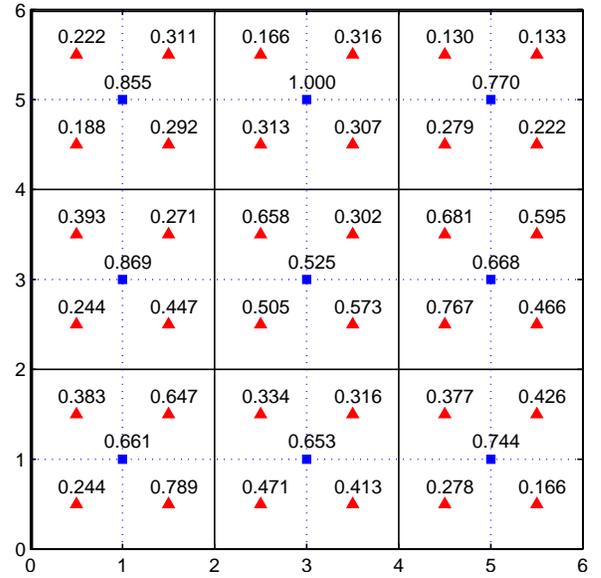}
\label{fig:demand_d05_LOG_load}}
\;\;
\subfigure[][Optimal demand allocation.]
{\includegraphics [scale=\Scale] {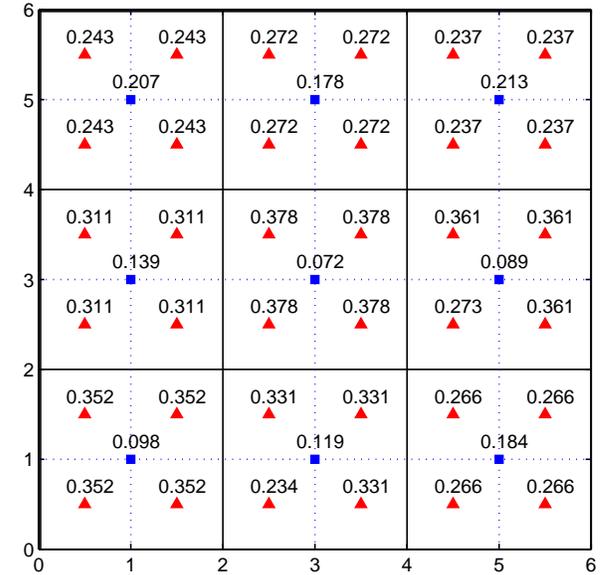}
\label{fig:demand_d05_LOG_demand}}
\caption{Optimal allocation with LOG utility and maximum demand fixed as $0.45$. The spectral radius is subject to a maximum constraint of $0.2694$ so that the maximum load value is less than one.} \label{fig:demand_d05_LOG}
\end{figure}

After constraining the load such that $\lmax(\rho^{\star})\leq 1$ via Problem~$P2$, we obtain the optimal demand vector $\bm{d}(\rho^{\star})$, and the corresponding load $\loadv^{\star}(\rho^{\star})$, as shown in Fig.~\ref{fig:demand_d05_LOG}. From Fig.~\ref{fig:demand_d05_LOG_load}, all the loads have been constrained to less than one.
Similar to the low maximum demand case, the load allocation is not uniform due to the non-uniform user allocation.
However, in contrast to the uniform demand allocation for the low maximum demand case shown in Fig.~\ref{fig:demand_d01_demand}, the demand allocation in Fig.~\ref{fig:demand_d05_LOG_demand} is not uniform. For example, in Fig.~\ref{fig:demand_d05_LOG_demand}, the base station at coordinates $(5,3)$ serves $0.089$~nat to all its users, while the WiFi access points within the base station cell serve a variation of demand, ranging from $0.273$~nat for the access point at $(4.5,2.5)$, to $0.361$~nat for the access points at $(4.5,3.5), (5.5,2.5)$ and $(5.5,3.5)$.
That is, the users are served in total a demand ranging from $0.362$~nat to the maximum request demand of $0.45$~nat.
The reason in the difference of the demand served is likely because the users that are served a smaller demand are closer to the center of the entire network and hence received more interference.
We note that this observation may not always hold in general since it depends on the user distribution and the resulting optimal load allocation in the entire network. For example, all the users served by the base station cell at $(3,3)$ receive the maximum demand, with $0.072$~nat from the base station and $0.378$~nat from their respective access points.
In general, however, we may still conclude that for the high maximum demand case, the spectral-radius constraints becomes tight which can limit the demand served to the users, especially those cells that receive the most amount of interference.

\begin{figure}
\centering
\subfigure[][Optimal load allocation.]
{\includegraphics [scale=\Scale] {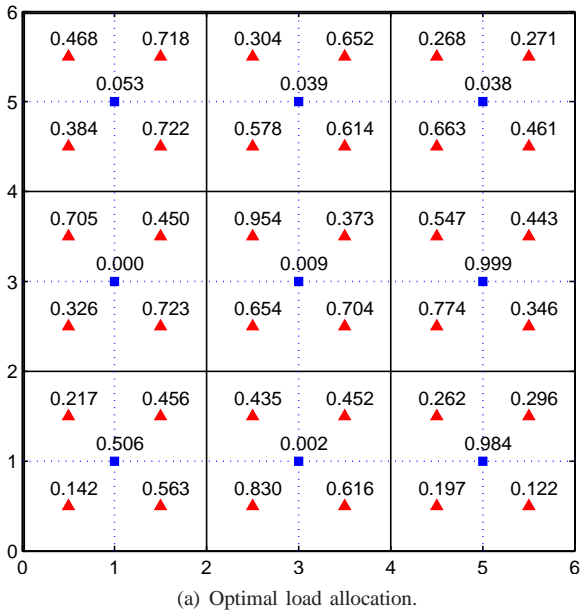}
\label{fig:demand_d05_LIN_load}}
\;\;
\subfigure[][Optimal demand allocation.]
{\includegraphics [scale=\Scale] {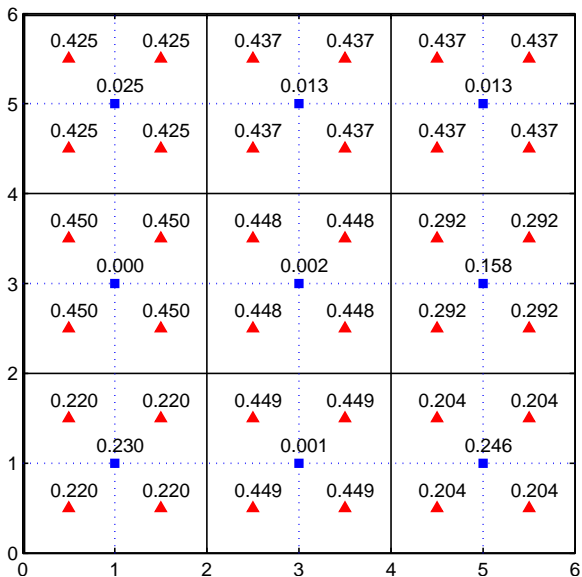}
\label{fig:demand_d05_LIN_demand}}
\caption{Optimal allocation with LIN utility and maximum demand fixed as $0.45$. The spectral radius is subject to a maximum constraint of $0.301$ so that the maximum load value is still less than one.} \label{fig:demand_d05_LIN}
\end{figure}

\subsection{Different Utility Functions}

We assume the high-maximum-demand case of $D_i=0.45, i\in \mathcal{N}$. The results for the LOG utility has been given earlier in Fig.~\ref{fig:demand_d05_LOG}.
The results for DLOG utility are almost identical to the case of the LOG utility, and are thus omitted. The similarity of the result is likely because the user fairness has already been largely taken into account via the LOG utility, and emphasizing this same aspect via the DLOG utility does not lead to a significantly different optimal solution.

Finally, we consider the use of the LIN utility. From Theorem~\ref{thm:general}, the feasible set $\widetilde{\mathcal{F}}$ may not be convex and hence our numerical solution is not necessarily optimal. Nevertheless, we shall see that we can still obtain a higher average per-user demand, but at the expense of user fairness.

%

The optimal demand vector $\bm{d}(\rho^{\star})$, and the corresponding load $\loadv^{\star}(\rho^{\star})$, are shown in Fig.~\ref{fig:demand_d05_LIN}. Again, we have constrained the load to be less than one, similarly by solving Problem~$P2$ as before.
The average per-user demand achieved is observed to be the maximally possible given by $0.45$~nat, compared to $0.4449$~nat achieved with LOG utility. This is within expectation since the LIN utility only focus on maximizing the sum demand. However, not all base stations or access points are uniformly served similar demand. In extreme cases, it is possible that some users are not served at all while other users are served the maximum load.

\section{Conclusion}\label{sec:con}

We have presented a utility-based optimization framework for data offloading in cellular networks, taking into account the inherent coupling relation among the cells.
Within this framework, fundamental properties on the computation, feasibility, and monotonicity of the load-coupling system have been studied. Three utility functions that differ in the emphasis on fairness have been considered, and fundamental insights of convexity analysis of the resulting optimization problem have been developed. Our analysis shows that optimal offloading is tractable when fairness is stressed.
We also propose a strategy to constrain the load to some maximum value, as required for practical implementation, and provide theoretical justification for the proposed algorithm.
In conclusion, our work provides a structured view on the offloading problem, and our analysis serves as a theoretical reference for empirical simulations and further performance evaluation.
As future work, we shall consider the related problems of energy minimization and user-network association.
\appendices

\section{Proof of Theorem~\ref{thm:d_inc_with_x}}\label{app:d_inc_with_x}
Consider the following asynchronous iteration in Remark~\ref{rem:asyn_itn} that runs for $k=1, \cdots, K$. For each {\em outer iteration} $k$, we execute an {\em inner iteration} that runs for $m=1,\cdots, n$:
\be\label{eqn:algoasyn}
\load^{k}_{m}= f_m(\loadv^{k-1,m})
\ee
where $\loadv^{0,1}=[\load^{0}_{1}, \cdots, \load^{0}_{n}]$ is an arbitrary initial load and $\loadv^{k-1,m}\triangleq[\load^{k}_1, \cdots, \load^{k}_{m-1}, \load^{k-1}_{m}, \cdots, \load^{k-1}_{n}]$ denotes the most current updated load vector.
After all iterations, the final load vector is given by $\loadv^{K,n+1}$, denoted simply as $\loadv^{K}$.
From Remark~\ref{rem:asyn_itn}, \eqref{eqn:algoasyn} is an asynchronous iteration which ensures that $\loadv^{K}$ converges to the final fixed-point solution in \eqref{eqn:nonlinearprob}. 

Suppose only one element of $\bm{d}'$ is strictly greater than $\bm{d}$, say $d_{ij}$. For the initial load, we choose $\loadv^{1,0}=\loadv^{\star}$. We then obtain the following results by performing the iteration \eqref{eqn:algoasyn}.
\begin{itemize}
\item For $k=1$: $x^1_i>x^0_i$ while $x^1_{\ell}=x^0_{\ell}$ for ${\ell}\neq i$.
\item For $k=2$: $x^2_i=x^1_i$ while $x^2_{\ell}>x^1_{\ell}$ for ${\ell}\neq i$.
\item For $k\geq 3$: $x^k_i>x^{k-1}_i$ while $x^{k}_{\ell}>x^{k-1}_{\ell}$ for ${\ell}\neq i$.
\end{itemize}
Specifically, for $k=1$, we have used \eqref{eqn:nonlinearprob0} where we replace $d_{ij}$ by $d_{ij}'$; for $k\geq 2$, we have used \eqref{eqn:algoasyn} and the result for the prior $k$.
Thus, $\loadv^{k,n}>\loadv^{k-1,n}$ for $k\geq 3$, while $\loadv^{2,n}\geq\loadv^{1,n}\geq\loadv^{1,0}=\loadv^{\star}$.
Together with the convergence guarantee, we get $\lim_{K\rightarrow\infty}\loadv^{K}_{n}={\loadv'}^{\star}>\loadv^{\star}$ as desired.
It is easy to check that the above conclusion holds even if more than one element in $\bm{d}'$ is strictly greater than $\bm{d}$ if all the users are served by the same (and only) base station $i$.

Next, consider the general case where $\bm{d}'\geq \bm{d}, \bm{d}'\neq \bm{d},$ Let $s\geq 1$ be the number of base stations serving users with different demand in $\bm{d}'$ and $\bm{d}$.
Then we can always find a set $\{\widetilde{\bm{d}}_1, \cdots, \widetilde{\bm{d}}_s\}$ with distinct elements ordered according to $\bm{d}'\geq \widetilde{\bm{d}}_s\geq \cdots \geq \widetilde{\bm{d}}_1\geq \bm{d}$ such that for any neighbouring pairs of vectors, e.g. $\{\bm{d}', \widetilde{\bm{d}}_s\}$, only one base station serve the users with different demand.
We then use the following inductive steps to complete the proof.
First, we obtain the load $\widetilde{\loadv}_1$ that corresponds to $\widetilde{\bm{d}}_1$ in \eqref{eqn:nonlinearprob}. To do so, we use $\loadv^{\star}$ as the initial load and the asynchronous iteration as before, which shows that $\widetilde{\loadv}_1>\loadv^{\star}$. Second, we use $\widetilde{\loadv}_1$ as the initial load and the asynchronous iteration as before, to show that the corresponding load $\widetilde{\loadv}_2$ satisfies $\widetilde{\loadv}_2>\widetilde{\loadv}_1$.
Proceeding similarly, we thus get ${\loadv'}^{\star}>\widetilde{\loadv}_s>\cdots>\widetilde{\loadv}_1>\loadv^{\star}$.

\section{Lemma on Irreducible Matrix}\label{app:irreducible}
Consider a non-negative matrix $\bm{B}\in \mathcal{R}_+^{n\times n}$ with the $(i,j)$th element given by $b_{ij}$. Let the {\em incidence matrix} of $\bm{B}$ be $\bm{A}\in \{0,1\}^{n\times n}$ with the $(i,j)$th element $a_{ij}= 1$ if $b_{ij}>0$ and $a_{ij}= 0$ if $b_{ij}=0$.
Denote the element of $\bm{A}^m$ as $a_{ij}^{(m)}$.
We say $\bm{B}$ is {\em irreducible} if $a_{ij}^{(m)}>0$ for all $i,j$ for some $m\geq 1$.

\begin{lemma}\label{lem:irreducible}
The matrix $\widetilde{\bm{\Lambda}}\in \mathcal{R}_+^{n\times n}, n\geq 3$, with $(i,k)$th element given by \eqref{eqn:def:lambdatilde} is irreducible. Also, $(\widetilde{\bm{\Lambda}}\widetilde{\bm{\Lambda}}^T)$ is irreducible.%
\end{lemma}
\begin{IEEEproof}
From \eqref{eqn:def:lambdatilde}, the diagonal elements of $\widetilde{\bm{\Lambda}}$ are zeros, while the off-diagonal elements are strictly positive since the channel gains $\{g_{kj}\}$ are positive. Thus, the incidence matrix of $\widetilde{\bm{\Lambda}}$ is $\bm{A}=\bm{1}_n\cdot \bm{1}_n^T - \bm{I}_n$, where $\bm{I}_n$ is the $n$-by-$n$ identity matrix. Thus $\bm{ A}^2=(n-2)\bm{1}_n\cdot \bm{1}_n^T + \bm{I}_n$.
Clearly $\bm{A}^2>{0}$, and so $\bm{\Lambda}(\bm{d})$ is irreducible.
Moreover, $(\widetilde{\bm{\Lambda}}\widetilde{\bm{\Lambda}}^T)$ is irreducible as the incidence matrix is $\bm{ A}^2>{0}$.
\end{IEEEproof}

\section{Lemma on Log-Convexity to prove Theorem~\ref{thm:general}}\label{app:logconvex}

\begin{lemma}\label{lem:logconvex}
Assume $f(x)$ is an increasing function with inverse $g(y)=f^{-1}(y)$.
Assume $f(x)$ and $g(y)$ are differentiable.
Then $g(y)$ is strictly log-convex if and only if
\be\label{eqn:lem:logcon2}
xf''(x)+f'(x) < 0.
\ee
\end{lemma}
where $f'(\cdot)$ and $f''(\cdot)$ are the first and second derivatives of $f(\cdot)$.
\begin{IEEEproof}
The second derivative of $\log(g(y))$ is given by $g''(y)/g(y)-(g'(y)/g(y))^2$, so $g(y)$ is strictly log-convex iff
\be\label{eqn:lem:logcon}
g''(y)g(y)- (g'(y))^2> 0.
\ee
To complete the proof, we shall show that \eqref{eqn:lem:logcon} holds.

We can write  $g(f(x))=x$.
Differentiating with respect to $x$, we get $g'(f(x))=1/f'(x)$.
Differentiating again with respect to $x$, we get $g''(f(x))=-f''(x)/(f'(x))^3$.
Thus the left-hand side of \eqref{eqn:lem:logcon} can be written as
$ g''(f(x)) g(f(x)) - (g'(f(x)))^2 =-x f''(x)/(f'(x))^3 - 1/(f'(x))^2 = -(x f''(x) + f'(x))/(f'(x))^3.$
Since $f'(x)\geq 0$, \eqref{eqn:lem:logcon} holds if and only if \eqref{eqn:lem:logcon2} holds, which completes the proof.
\end{IEEEproof}

\section{Lemma to Prove Theorem~\ref{thm:U_inc}}\label{app:r_inequality}

\begin{lemma}\label{lem:r_inequality}
Let $\bm{A}, \bm{B}\geq 0$ be $n$-by-$n$ matrices, and $\bm{A}$ is irreducible. Then $r(\bm{A}+\bm{B})\geq r(\bm{A})$ with equality if and only if $\bf{B}=0$.
\end{lemma}
\begin{IEEEproof}
Let $\bm{u}$ be the (right) eigenvector that corresponds to the largest eigenvalue of $\bm{A}$.
Applying the Perron-Frobenius Theorem to $\bm{A}$ \cite{Stanczak09}, we have $\bm{u}> 0$ and $r(\bm{A})$ equals the largest eigenvalue.
Then the spectral radius of $\bm{A}+\bm{B}$ can be written as
\be
r(\bm{A}+\bm{B})&=& \max_{\|\bm{z}\|=1} | \bm{z}^H(\bm{A}+\bm{B})\bm{z} | \\
&\geq& \bm{u}^H\bm{A}\bm{u}+\bm{u}^H\bm{B} \bm{u} \\
&\geq& \bm{u}^H \bm{A}\bm{u}  = r(\bm{A})
\ee
Here, the first inequality is due to replacing $\bm{z}$ by $\bm{u}$. The second inequality is due to $\bm{B}\geq 0$ and $\bm{u}> 0$, and becomes an equality if and only if $\bf{B}=0$. This concludes the proof.
\end{IEEEproof}


\end{document}